\documentclass[showpacs,pra,aps,twocolumn]{revtex4}
\usepackage{graphicx}
\usepackage{psfrag}
\usepackage{amsmath,amssymb}
\usepackage{colordvi}

\renewcommand{\a}{\alpha}
\renewcommand{\b}{\beta}
\newcommand{\g}{\gamma}

\renewcommand{\d}{\delta}
\renewcommand{\S}{\Sigma}
\newcommand{\s}{\sigma}
\newcommand{\D}{\Delta}

\renewcommand{\o}{\omega}
\renewcommand{\O}{\Omega}
\newcommand{\e}{\epsilon}
\renewcommand{\l}{\lambda}

\renewcommand{\dag}{\dagger}
\newcommand{\lb}{\label}
\newcommand{\nn}{\nonumber}

\newcommand{\PD}[2]{\frac{\partial{#1}}{\partial{#2}}}

\newcommand{\bk}[1]{\left(#1\right)}
\newcommand{\bra}[1]{\langle{#1}|}
\newcommand{\ket}[1]{|{#1}\rangle}
\newcommand{\lr}[1]{\left\langle#1\right\rangle}

\newcommand{\be}{\begin{equation}}
\newcommand{\ee}{\end{equation}}
\newcommand{\ba}{\begin{eqnarray}}
\newcommand{\ea}{\end{eqnarray}}

\begin{document}

\title{Universal Properties of the Ultra-Cold Fermi Gas}
\author{Shizhong Zhang and Anthony J. Leggett}
\affiliation{Department of Physics, University of Illinois at
Urbana-Champaign, 1110 West Green Street, Urbana, Illinois, 61801-3080.}
\date{\today}
\begin{abstract}
We present some general considerations on the properties of a
two-component ultra-cold Fermi gas along the BEC-BCS crossover. It is
shown that the interaction energy and the ground state energy can be
written in terms of a single dimensionless function $h({\xi,\tau})$,
where $\xi=-(k_Fa_s)^{-1}$ and $\tau=T/T_F$. The function $h(\xi,\tau)$
incorporates all the many-body physics and naturally occurs in other
physical quantities as well. In particular, we show that the
RF-spectroscopy shift $\overline{\d\o}(\xi,\tau)$ and the molecular
fraction $f_c(\xi,\tau)$ in the closed channel can be expressed in
terms of $h(\xi,\tau)$ and thus have identical temperature
dependence. The conclusions should have testable consequences in future
experiments.  
\end{abstract}

\maketitle

\section{Introduction}
Over the past few decades, there have been considerable efforts and
progress in understanding the physics of ultra-cold Fermi 
gases\cite{Randeria1995,Bloch2008,Pita2008,Ketterle2007}. In general, the
theoretical investigations fall into two 
categories depending on how one incorporates the physics of Feshbach
resonances. In the so-called single-channel
model, one
neglects the closed channel component, while incorporating its effects
through the open channel scattering length $a_s$, given by,
\ba
a_s(B)=a_{bg}\bk{1-\frac{\D B}{B-B_0}}
\ea 
where $a_{bg}$ is the background scattering length in the absence of
inter-channel coupling and $B$ is the external magnetic field. $\D B$
is the width of the resonance and $B_0$ is the position of the
resonance. See Table.V. in Ref.\cite{Kohler2006} for a list of values of
these parameters for the alkali elements currently under
investigation. The approximation is valid in the case of a  so-called
broad resonance
where $\e_F\ll \d_c$. Here $\e_F$ is the Fermi energy of the system
and $\d_c=\frac{(\D\mu \D B)^2}{2\hbar^2/ma_{bg}^2}$\cite{note-dc},
characterizing the  
typical energy scale associated with the two-body Feshbach
resonance\cite{Leggett2006}. 
$\D\mu$ is the magnetic moment difference between open and closed
channels. The physics of the single channel model is essentially the
same as in  
the crossover model studied decades ago in the
literature\cite{Eagles1969,Leggett1980,Nozieres1985}. Most
experimental systems ({\it e.g.} $^6$Li at 
magnetic field $B=834G$) fall into this category. In
general, any system with sufficiently low density will be described by
a single channel model.

The problem associated with single-channel model is easily
stated: Given a spin-1/2 Fermi gas with both spin
components equally populated,
with interactions only between opposite spin states,
characterized by the s-wave scattering length $a_s(B)$, what are
the ground state and thermodynamic properties? At zero temperature,
neglecting finite range corrections, the only relevant parameter is
$\xi=-(k_Fa_s)^{-1}$. The basic question concerning the static
properties of the system  is how to find the ground state energy 
$E(\xi)$.  At finite temperature, the most important questions are the
calculation of the thermodynamic potential and in particular the
location of the 
phase transition boundary $T_C(\xi)$ as a function of
$\xi$; see Ref.\cite{Haussmann2007} and references therein. Up to now,
these problems have not been amenable to analytic solutions; our most
reliable knowledge of those quantities comes from Monte Carlo
simulations\cite{Carlson2003,Astrakharchik2004,Chang2005,Burovski2006,
  Bulgac2007,Akkineni2007,Gezerlis2008}.   

For most experiments, it is sufficient to use the single-channel model
as the theoretical framework to interpret them. However, there are
certain cases where we are interest in the physics of the closed channel
specifically for a broad Feshbach resonance, as in the case of the Rice
experiment to be discussed later\cite{Partridge2005}.  Furthermore, in
the so-called narrow resonance case where 
$\e_F\gtrsim\d_c$, the molecular states in the closed
channel cannot be neglected. For these reasons, we have to start 
with the more general 
two-channel model\cite{Holland2001,Timmermans2001, Levin2005}, which
incorporates the closed channel on 
the same footing as the open channel. Typically, one introduces a
bosonic operator $\Psi({\vec r})$, which creates a closed channel
molecule with 
center of mass position ${\vec r}$ and incorporates its effects
through the coupling term in the Hamiltonian,
\ba
\tilde{g}\bk{\Psi^\dag(\vec{r})\psi_{\uparrow}(\vec
  r)\psi_{\downarrow}(\vec r)+\mbox{H.C.}}, 
\ea
where $\tilde{g}$ is the bare coupling constant. Note that the coupling
scheme above enforces the momentum conservation in the conversion
processes and thus the momentum distribution of closed channel
molecules is intimately connected with the open channel pair
states. We shall return to this point later. Note also that the
internal structure of the closed channel molecule is frozen as a
result of 
its high internal excitation energy which is much larger than any
other energy scales relevant for the many-body physics.

In general, exact solutions or even general statements about the above
two models are quite difficult. However, at resonance, {\it i.e.} $
\xi=0$, $a_s$ drops out of the problem and we are left with only two
energy scales, $\e_F=\hbar^2/2m(3\pi^2n)^{2/3}$ and the temperature
$T$. Here $m$ is the mass of the atom and $n$ is the density of 
the system. Thus, it follows from dimensional analysis that one may
write the average single particle energy of the system at finite
temperature as $\e(\xi=0,\tau)=\e_F f_E(\xi=0,\tau)$, where 
$f_E(\xi=0,\tau)$ is a dimensionless function and $\tau=T/T_F\equiv
k_BT/\e_F$, where $k_B$ is the Boltzmann constant. In particular, at
$\tau=0$, the average single particle  energy is proportional to
$\e_F$, with universal constant $f_E(\xi=0,\tau=0)\equiv
\frac{3}{5}(1+\b)$. The parameter  
$\b$ has been calculated in many ways in the literature and is in good
agreement with experiments. See Table.II. in \cite{Pita2008} for a
summary of values of $\b$ obtained theoretically and experimentally. By
the same argument 
one can write down other thermodynamic quantities of the system, 
with the conclusion that the thermodynamic properties of the system are
universal regardless of the particular system under
investigation\cite{Ho2004}.

The universal thermodynamics works only at unitarity. However, it is
possible to generalize the idea to the parameter space where $\xi\neq
0$. It is recognized that for $d>2$, the system is controlled by an
unstable fixed point which resides near the Feshbach resonance in the
zero density limit with 
attractive interactions\cite{Nikolic2007}. By utilizing a large-N
expansion as applied 
to Sp(2N) model, one can calculate the scaling form of the canonical
free energy of the system. The $\e$-expansion has also been used to
investigate the properties of the two-component Fermi gas away from
resonance\cite{Nishida2007,Chen2007}. 

There is however, another form of  `universality' which is more
deeply rooted in the actual physical properties of the system. For
example, by examining the short-range form of the many-body wave
function, one can show that several physical quantities(interaction
energy, RF-spectroscopy shift {\it etc.}) depend on temperature
$\tau$ through one universal function $h(\xi,\tau)$. This universal
dependence comes about because of one peculiar property of the dilute
Fermi gases: the range of the interaction is much smaller than
the inter-particle distance, {\it i.e.} $k_F r_0\ll 1$, where $r_0$ is
the range of the potential. Thus, important effects associated with
interactions come mostly from two-body encounters. The argument
presented below can then be regarded as an expansion in
terms of $k_F r_0$. Let us note that this argument can be trivially
modified in the case of the imbalanced Fermi gas.

The organization of the paper is the following. In Sec.II, we give a
general discussion of the physical system in terms of the two-body density
matrix and separate the two-body and many-body contributions in it. In
Sec.III, we apply the result of Sec.II to several physical quantities
and show that they can be written in terms of one universal function
$h(\xi,\tau)$ which carries all the many-body dependence. There are
residual $\xi$-dependences as a result of the two-body physics, which
can in principle be calculated without any reference to the many-body
system. The temperature dependence of those physical quantities is
universal and has experimental consequences as described in
Sec.III. In Sec.IV, the main conclusions of the paper are
summarized and discussed. In the Appendix, we give an alternative
derivation of the linear $\xi$-dependence of the energy of the system
away from resonance, based on the many-body wave functions.

\section{General Setup}
The difficulties involved in analyzing either the single-channel or the
two-channel model are often expressed as a lack of small parameter
because of the resonant interaction condition $na_s^3\gg 1$, which
prevents a relatively 
straight-forward perturbation calculation as in the classic dilute
Fermi gas\cite{Abrikosov1975,Huang1957,Lee1957}. The challenge lies in
the correct implementation of the 
two-body physics, characterized by the `dangerous' diverging
scattering length $a_s$, into the many-body calculations. One way to
circumvent the difficulty is to devise other small parameter, as in
the $\e$-expansion\cite{Nishida2007} or $1/N$-expansion\cite{Nikolic2007}
utilized in recent work. On the 
other hand, even though the parameter $na_s^3 \gg 
1$, we still have the small parameter $k_Fr_0$, where $r_0$ is the
range of the potential. Note that in most investigations using the
single- or two-channel model, the zero-range limit has already been
taken; an exception is the case of a narrow
resonance\cite{Gurarie2007}. In the 
following, we will try to set up an approximation scheme which utilizes
the smallness of $k_Fr_0$. Even though it does not yield 
immediately a computational tool for the values of specific
constants, say $\b$, it does lead to some general conclusions {\it
  independent} of the approximation scheme employed in a specific
investigation.

 To motivate our discussion, let us consider the
many-body wave function for a spin-1/2 Fermi system with both spin
components equally populated. We denote the
 total number of atoms $N$. For example, we consider a collection of
$^6$Li atoms in their lowest two hyperfine-Zeeman states($\ket{1}$ and
$\ket{2}$).  Let us 
write down its many-body wave function as $\Psi({\vec r}_1\s_1,{\vec
  r}_2\s_2 \cdots {\vec r}_N\s_N)$. Now, we separate two atoms, say
atom 1 and atom 2, with opposite spin orientations, far (compared to
$r_0$) from all the other atoms ($3,4,\cdots ,N$), and ask what is the form
of the many-body wave function, when the distance 
between ${\vec r}_1$ and ${\vec r}_2$ is taken to lie within the range
of two-body interacting potential, {\it i.e.}, $|{\vec r}_1-{\vec r}_2|
\lesssim r_0$. Since all the other $N-2$ atoms cease to interact with
atom 1 and atom 2, we conclude that
\ba
&&\lim_{|{\vec r}_1-{\vec r}_2| \lesssim r_0} \Psi({\vec r}_1\s_1,{\vec
  r}_2\s_2 \cdots {\vec r}_N\s_N)\\\nn
&\propto& \mathcal{A} \phi({\vec
  r}_1-{\vec r}_2)\S_{12}\Psi'({\vec r}_3\s_3,{\vec
  r}_4\s_4 \cdots {\vec r}_N\s_N)  
\ea
Here $\mathcal{A}$ is the trivial anti-symmetrization operator and 
$\S_{12}$ is the spin wave function of atoms 1 and 
2. $\phi({\vec r}_1-{\vec r}_2)$ is determined by the two-body
interaction potential in the range  $|{\vec r}_1-{\vec r}_2|
\lesssim r_0$. In order for the other $N-2$ atoms to affect
the form of $\phi({\vec r}_1-{\vec r}_2)$, it is necessary that a third
atom is at a distance $\lesssim r_0$. Such process is highly
unlikely for two 
reasons. Firstly, the phase space of such event is down by at least a
factor $(k_F r_0)^3$ as compared with the two-body encounters. Secondly,
in a spin-1/2 system, two of the three atoms close together must have
the same spin, thus Pauli exclusion principle will prevent such
process from 
occurring. In fact, to the extent that we can work entirely in terms of
the s-wave scattering length $a_s$, we have automatically neglected the
contributions from higher partial wave scattering, which are of order
$(k_Fr_0)^3$ or higher. We thus conclude that, to order $k_F r_0$, the
short-range behavior 
of the many-body wave function(in particular its nodal structure) is
determined by two-body physics. An important question pertains to the
form of two-body wave function $\phi({\vec r}_1-{\vec r}_2)$, since in
general, the two-body potential can host several bound states. Here we
note that since the many-body energy scale is much smaller than the
energy splitting between the different energy levels in the potential
well, it is easy to convince oneself that only the bound state that is
closest to the scattering continuum is relevant. This state is nothing
but the molecular state on the BEC side of the crossover. Here we
emphasize that even in the BCS-side, where the two-body bound state
emerges above the scattering continuum, the above conclusion
still holds. To conclude this intuitive discussion,
we must point out that at this stage, we do not yet know the
normalization of the short-range part of the many-body wave
functions. This will be determined by the many-body physics.

To make the above argument more precise, let us consider the two-body
density matrix for a generic many-body system. The definition of the 
two-body density matrix is given by\cite{Leggett2006} 
\ba\label{dm}
&&\rho({\vec r}_1\a,{\vec r}_2\b;{\vec r}_3\g,{\vec r}_4\d;t)\\\nn
&=&\lr{\psi^\dag_\a({\vec r}_1,t)\psi^\dag_\b({\vec r}_2,t)
  \psi_\g({\vec r}_3,t)\psi_\d({\vec r}_4,t)}.      
\ea
Here $\psi_\a({\vec r}_1,t)$ is the Heisenberg field operator for a 
fermion 
with spin $\a$. In the following, we shall consider only an equilibrium
situation and thus drop the time $t$ from the above expression. By the 
Hermicity property of the density matrix, we can decompose 
the two-body density matrix in the following form\cite{Leggett2006}, 
\ba
&&\rho({\vec r}_1\a,{\vec r}_2\b;{\vec r}_3\g,{\vec r}_4\d)\\\nn
&=&\sum_in_i\phi^{(i)*}_{\b\a}({\vec r}_2,{\vec r}_1) 
\phi^{(i)}_{\g\d}({\vec r}_3,{\vec r}_4). 
\ea
The eigenvalues $n_i$ and eigenfunctions
$\phi_{\a\b}^{(i)}(\vec{r}_1,\vec{r}_2)$ 
satisfy the following conditions, 
$\sum_in_i=N(N-1)$ and $\sum_{\a\b}\int d^3{\vec r}_1\int
d^3{\vec r}_2\phi_{\a\b}^{(i)*}({\vec r}_2,{\vec
  r}_1)\phi_{\b\a}^{(j)}({\vec r}_1,{\vec r}_2)=\d_{ij}$. 
As discussed above, in the case of a dilute Fermi gas, the only relevant
parameter is $\xi=-(k_Fa_s)^{-1}$ and $\tau$. Thus $n_i$ and
$\phi_{\b\a}({\vec r}_2,{\vec r}_1)$ will depend on $\xi$ and $\tau$
parametrically. Now, by the argument given above in terms of the many-body
wave function, we see that the
short-range form of $\phi_{\a\b}({\vec r}_1-{\vec r}_2)$ will be
determined by two-body 
physics, while many-body physics will determine the eigenvalues $n_i$
and the long range part of the eigenfunctions. Our philosophy in the
following will be to express several physical 
quantities in terms of the two-body density matrix and use the above facts
to extract their universal dependence on temperature. To be
successful, we need our expressions to pick up only the 
short range part of the two-body density matrix so that all the
temperature dependence will be carried by
$n_i$'s and the normalizations for the pair wave
functions. Physically, as we change the temperature and
the interaction strength, the occupation numbers of the pair wave functions
$\phi_{\a\b}^{(i)}$ change while the short range part of the pair
wave function remains the same. Let us thus consider an arbitrary
short-range($\sim r_0$) 
function $s({\vec r}_1-{\vec r}_2)$ and consider the integral,
\ba\label{samp}\nn
&&\int d{\vec r}_1 d{\vec r}_2 s({\vec r}_1-{\vec r}_2)\lr{\psi_1^\dag({\vec
    r}_1)\psi_2^\dag({\vec r}_2)\psi_2({\vec r}_2)\psi_1({\vec
    r}_1)}\\
&=&\sum_i n_i(\xi,\tau) \int s({\vec r}_1-{\vec r}_2)|\phi^{(i)}_{12}({\vec
  r}_1,{\vec r}_2)|^2 d{\vec r}_1 d{\vec r}_2.
\ea
It is clear that in the above equation, we need only retain the s-wave
part of the pair wave function, since higher partial waves have
vanishing probability at the origin and thus hardly contribute to the
above integral. Thus, we can write $\phi_{12}^{(i)}({\vec r}_1,{\vec
  r}_2)= \O^{-1/2}e^{i{\vec
  p}\cdot\frac{\vec{r}_1+\vec{r}_2}{2}}\chi_{12}^{(i)}(r)Y_{00}/r$,
where $r=|{\vec r}_1-{\vec r}_2|$ and $Y_{lm}$ is the $l=m=0$ spherical
harmonics. The factor $\O^{-1/2}e^{i{\vec
  p}\cdot\frac{\vec{r}_1+\vec{r}_2}{2}}$ describes the center of mass motion of
the pair state. Then the R.H.S. of Eqn.(\ref{samp}) can be written as, 
\ba
\sum_i n_i(\xi,\tau) \int dr s(r)|\chi^{(i)}_{12}(r)|^2. 
\ea  
By the above argument,
$\chi_{12}(r)$ will have the form of the two-body radial wave
function at short distance. In particular, in the region where
$a_s,k_F^{-1} \gg r \gtrsim r_0$, we can write
$\chi_{12}(r)=C^{(i)}(\xi,\tau)\overline{\chi}^{(i)}_{12}(r)$, where,   
\ba
\overline{\chi}^{(i)}_{12}(r)\equiv 1-\frac{r}{a_s}
\ea
has been normalized in such a way that
it approaches $1$ in the region $a_s,k_F^{-1} \gg r \gtrsim
r_0$. The $C^{(i)}$'s are some constants which in principle depend on
the many-body physics. Eqn.(\ref{samp}) can then be written as,
\ba\label{defh}\nn
&&\int d{\vec r}_1 d{\vec r}_2 s({\vec r}_1-{\vec r}_2)\lr{\psi_1^\dag({\vec
    r}_1)\psi_2^\dag({\vec r}_2)\psi_2({\vec r}_2)\psi_1({\vec
    r}_1)}\\\nn
&=&\sum n^{(i)}(\xi,\tau) |C^{(i)}(\xi,\tau)|^2 \int dr
s(r)|\overline{\chi}_{12}(r)|^2\\
 &\equiv& h(\xi,\tau)k_F N\int dr
s(r)|\overline{\chi}_{12}(r)|^2, 
\ea
where we have defined a {\it positive definite} universal function
\ba
h(\xi,\tau)\equiv \sum_i\frac{n^{(i)}(\xi,\tau)}{Nk_F}
|C^{(i)}(\xi,\tau)|^2>0.
\ea
The factor $k_F$ is inserted in order to make $h(\xi,\tau)$ a
dimensionless function. The integral in
Eqn.({\ref{defh}}) is a constant depending on the function $s(r)$, but
it is purely a two-body quantity and can be calculated without making
reference to the many-body system. In particular, it  does {\it not}
depend on the temperature $T$. We further note that the integral
displays no singular dependence on $a_s$ as we approaches the
resonance. Thus for the discussion of many-body physics, it can be
regarded as a known parameter. The intricate many-body correlations
are then incorporated in one {\it universal} function $h(\xi,\tau)$
and are themselves universal. As we shall
show later, at unitarity, $h(\xi=0,\tau)$ must be finite and thus we
conclude Eq.(\ref{defh}) scales with $k_F$ at unitarity. 

Before ending the discussion of this section, let us remind ourselves
of the assumptions made so far:\\
\indent
$\a$), Only s-wave scattering is important. The neglect of higher
angular momentum($\hbar l$) partial waves is justified because there are of
relative order 
$(k_Fr_0)^{2l}$ and thus negligible as compared with s-wave
scattering. In fact, in the model Hamiltonian considered in the
literature, only s-wave scattering is included.\\
\indent
$\b$), The short-range form of the pair function
$\chi(\vec{r}_1-\vec{r}_2)$ is determined by two-body physics and
moreover, corresponds to {\it only} one particular form of the two-body
wave function in the range $\sim r_0$. The former assumption is
justified because of 1), the diluteness of the system $k_Fr_0\ll 1$
and 2), the `exchange hole': the Pauli principle forbids two particles
with like spin to be close to each other. The later assumption come
from 
energetic considerations: as long as we are interested in the many-body
physics, which has a typical energy scale $\e_F$, the relevant
two-body state is the one that is closest to the zero-energy
scattering state, with all the other two-body states far away to be of
any practical importance.


In the following, we shall consider a uniform system with density
$n$ at temperature $T$. The interactions between particles can be
written as
\ba
\frac{1}{2}\sum_{i,j}\bk{f(\vec{r}_i-\vec{r}_j)+g(\vec{r}_i-\vec{r}_j)
  \vec{S}_i\cdot \vec{S}_j}.
\ea
Here $f(\vec{r})$ and $g(\vec{r})$ are the direct and exchange
interaction respectively. $\vec{S}$ is the spin operator of the
valence electron of the atom under 
consideration. Experimentally, one normally works with an equal
population of atoms(say $^6$Li) in the lowest two hyperfine states
$\ket{1}$ and $\ket{2}$. To 
the extent that one can neglect the closed channel component, as is
the case for a broad resonance, one may replace the full interaction by
an effective short-range interaction in the open channel 
$V_\l(\vec{r})\equiv V({\vec r},\l)$, 
where $\l$ is a controlling parameter by which one can tune the 
scattering length $a_s$\cite{Leggett2006}. However, in discussing the
closed channel 
population, it is necessary to introduce explicitly the inter-channel
coupling term $W(\vec{r})$ which 
converts open channel pair states to closed channel molecules, see the
discussion in Sec.III.D.

\section{Physical Quantities}
In the following we shall discuss several physical quantities that can
be expressed in terms of the universal function $h(\xi,\tau)$ and thus
display a universal dependence on the temperature $T$.\\

\subsection{Interaction Energy}
The simplest physical quantity that can be cast in the form of
Eq.(\ref{defh}) is the interaction energy of the system. According to
the discussion above, since the 
interactions between particles are of short-range form, we can write
the interaction energy per particle $\frac{\lr{V}}{N}$ as,
\ba\lb{ie}\nn
&&\frac{1}{N}\int d{\vec r}_1d{\vec r}_2 V({\vec r}_1-{\vec
  r}_2)\lr{\psi_1^\dag({\vec r}_1)\psi_2^\dag({\vec r}_2)\psi_2({\vec
    r}_2)\psi_1({\vec r}_1)}\\
&=&C_V(a_s)k_F h(\xi,\tau),
\ea
where $C_V(a_s)\equiv\int dr V(r)|\overline{\chi}_{12}(r)|^2$ is a
well-defined, purely
two-body quantity. All the many-body dependence of interaction energy
is encapsulated 
in the universal function $h(\xi,\tau)$. Since the interaction energy
must be well-defined at unitarity, $h(\xi,\tau)$ must be
finite, and moreover free of any divergence as $\xi$ approaches
zero. Thus the average interaction energy 
per particle $\frac{\lr{V}}{N}$ scales as $k_F$ at unitarity. The
interaction energy of 
the system depends on microscopic details of the system, even at
unitarity, as is clear
from the factor $C_V(a_s)$. This result should be
compared with the total energy of the system, to be discussed in the
next subsection, which is proportional to the Fermi energy $\e_F$ at
unitarity, independent of microscopic details.

\subsection{Total Energy} 
To derive an expression for the total energy of the system, we
first recall that, if $\l$ is the tuning parameter of the potential by
which the scattering length can be varied(see Appendix), then the
relation between $a_s$ and $\l$ is given by,
\ba\lb{asl}
\d a_s^{-1}=-\frac{m}{\hbar^2}\bk{\int_0^\infty dr
  \PD{V(r,\l)}{\l}|\overline{\chi}_{12}(r)|^2} \d \l. 
\ea 
On the other hand, according to the Hellmann-Feynman theorem, we have, 
\ba\lb{fh}
\PD{E}{\l}=\lr{\PD{V({\vec r},\l)}{\l}},
\ea  
with the average taken over the many-body state as in (\ref{samp}). 
Since $\PD{V}{\l}$ is a short-ranged function, we can use
Eqn.(\ref{asl}) to rewrite Eqn.(\ref{fh}) in terms of $a_s^{-1}$; we
find,
\ba
\PD{E}{a_s^{-1}}=-\frac{\hbar^2}{m}N k_F h(\xi,\tau)
\ea
Here we have used the definition of $h(\xi,\tau)$ in
Eq.(\ref{defh}). Or in terms of $\xi=-(k_Fa_s)^{-1}$, 
\ba\lb{fh2}
\PD{E}{\xi}=\frac{\hbar^2k^2_F}{m}N h(\xi,\tau)=2\e_F N h(\xi,\tau)
\ea
Since by definition $h(\xi,\tau)$ is a positive definite function,
we find the somewhat trivial result that the ground state energy is a
{\it monotonically} increasing function of $\xi$. The boundary
condition on the above differential equation is easily
obtained. Consider the case when 
$\xi= +\infty$ and $\tau=0$, we then have a free Fermi gas with the
average single particle energy $\e(\xi=\infty)\equiv
\frac{E}{N}=\frac{3}{5}\e_F$.  Integrating Eqn.(\ref{fh2}),
we find that the single particle energy at zero temperature along the
BEC-BCS crossover is given by 
\ba\lb{spe}
\e(\xi)=\frac{3}{5}\e_F-2\e_F\int_\xi^\infty
h(\xi')d\xi',
\ea
where $\e(\xi)\equiv \e(\xi,\tau=0)$ and $h(\xi)\equiv
h(\xi,\tau=0)$. Intuitively, $h(\xi)$ accounts 
for the reduction of the single 
particle energy due to interaction effects. At unitarity
$\e_0\equiv \e(\xi=0,\tau=0)=(1+\b)\frac{3}{5}\e_F$, so we find 
\ba\lb{beta}
\b=-\frac{10}{3}\int_0^\infty h(\xi')d\xi' 
\ea
The generalization of the above expression to finite temperature is
straightforward but may be less useful. Around unitarity where
$\xi\ll 1$, we can obtain an expansion of the average single particle
energy at finite temperature in term of $\xi$. To this end, we can
integrate Eq.(\ref{fh2}) from 
$\xi'=0$ to $\xi'=\xi$ and we find,
\ba
\e(\xi,\tau)-\e(\xi=0,\tau)=2\e_F\int_0^\xi h(\xi',\tau)d\xi'
\ea 
For $\xi$ close to zero, the question reduces to the expansion of
$h(\xi)$. From the discussion in the previous subsection, we know
$h(\xi=0,\tau)$ is finite, so we conclude the energy correction away
from unitarity is linear in $\xi$ and given by,
\ba
\e(\xi,\tau)-\e(\xi=0,\tau)=2\e_F h(\xi=0,\tau)\xi+\cdots
\ea
At zero temperature, the value of $h(\xi=0,\tau=0)$ can be calculated
using the $\e$-expansion, 
where one has to sum over all the higher order logarithms in order to
recover the correct linear $\xi$-dependence of the
energy\cite{Chen2007}. Instead of 
referring back to the conditions imposed by the interaction energy  on the
function $h(\xi,\tau)$, one can have a direct derivation of the linear
$\xi$-dependence of the energy away from unitarity by a straightforward
generalization of the argument in the two-body case. This is presented
in the Appendix. 

Before we conclude this subsection, we would like to derive a simple
relation between the chemical 
 potential $\mu$ and the average single particle energy $\e$ at zero 
temperature and thus  
enable us to write down the zero temperature chemical potential in
terms of $h(\xi)$. Note that at $T=\tau=0$, we can write the
single particle energy as $\e=\e_Ff_E(\xi)$. Thus using the
thermodynamic relation $P=-\PD{E}{V}$ and $E=N\e$, we find,
\ba\lb{pressure}
p=n^2\PD{\e}{n}=n^2\PD{[\e_Ff(\xi)]}{n}.
\ea
Using the relation $n=\frac{k_F^3}{3\pi^2}$, we can make a change of
variable to $k_F$ and write Eq.(\ref{pressure}) as
$p=\frac{1}{3}nk_F\PD{\e}{k_F}$.  Now, using the expression
$\e=\e_Ff(\xi)$ and the fact that $f(\xi)$ only depends on the combination
$\xi\equiv -(k_Fa_s)^{-1}$, we obtain,
$p=\frac{2}{3}n\e+\frac{1}{3}na_s\PD{\e}{a_s}$. We consider the 
situation when the density of the system is fixed 
and write the above expression as
$\e=\frac{3}{2}\frac{p}{n}+\frac{1}{2}\xi\PD{\e}{\xi}$. At $T=0$, we 
have the thermodynamic relation, $\e=-\frac{p}{n}+\mu$, where $p$ is the
pressure and $n$ is the average density. We find,
\ba\lb{emu}
\frac{5}{2}\e=\frac{3}{2}\mu+\frac{1}{2}\xi\PD{\e}{\xi}.
\ea
The above expression is very general and works along the whole BEC-BCS
crossover provided that the density $n$ is kept constant. In the
extreme BEC limit, the single particle energy equals the
chemical potential: $\e=\mu=-\frac{\hbar^2}{2ma_s^2}$. One verifies
that this is satisfied 
by Eq.(\ref{emu}). In the BCS limit, the first order correction to
the energy will be of order $a_s$, namely of order $\xi^{-1}$, coming
from Hartree-Fock corrections. Since
$\xi\PD{\e}{\xi}=-\xi^{-1}\PD{\e}{\xi^{-1}}=-a_s\PD{\e}{a_s}\to 0$
when $a_s\to 0$, we thus recover the usual relation between chemical
potential $\mu$ and the average single particle energy of the free
Fermi gas $\e=\frac{3}{5}\mu$. At unitarity, $\xi=0$, we find again
the free Fermi gas result $\e=\frac{3}{5}\mu$, if we assume that the
energy is continuous at unitarity.

Finally, using Eqn.(\ref{fh2}), we can write the zero temperature
chemical potential in terms of $h(\xi)$ as,
\ba\lb{mu}
\mu(\xi)=\e_F\bk{1-\frac{2}{3}\xi h(\xi)-\frac{10}{3}\int_\xi^\infty
  h(\xi')d\xi'}. 
\ea
Setting $\xi=0$, we recover Eq.(\ref{beta}) since at unitarity
$\mu=(1+\b)\e_F$.

\subsection{RF-spectroscopy Shift $\overline{\d\o}$}
One of the early experiments which indicated the appearance of a new
low temperature quantum state in ultra-cold Fermi gases was 
the radio-frequency spectroscopic experiment carried out by the
Innsbruck Group\cite{Chin2004}. The experiment works
with the lowest two hyperfine-Zeeman states($\ket{1}$ and $\ket{2}$)
of $^6$Li. A radio-frequency field is applied to drive atoms from
state $\ket{2}$ to $\ket{3}$. It is found that at high temperature,
the frequency of the rf-field coincides with the bare atomic
transition from $\ket{2}$ to $\ket{3}$, while at low temperature,
there is an up-shift in the rf-frequency which indicates that the system
is in a new quantum state. It is now understood that the full
understanding of the RF-spectroscopic profile is quite complicated,
requiring a proper treatment of the final-state
interactions\cite{Kinnunen2004,Ohashi2005,He2005,Yu2006,Baym2007,
Punk2007,Perali2008,Zhang2008}.  It has been shown that the average
shift in the RF-spectroscopy is given by the following
expression\cite{Zhang2008}, 
\ba\lb{rf}\nn
\overline{\d\o}&=&\frac{G(H)+J(H)}{\hbar N_2}\\\nn
&&\times\int
g(\vec{r}_1-\vec{r}_2)\lr{\psi_1^\dag(\vec{r}_1)\psi_2^\dag(\vec{r}_2) 
  \psi_2(\vec{r}_2)\psi_1(\vec{r}_1)}\\
&=&\frac{G(H)+J(H)}{\hbar}C_g(a_s)2 k_F
h(\xi,\tau)
\ea
where $C_g(a_s)=\int g(r)|\overline{\chi}_{12}(r)|^2 dr$ and the
functions $G(H)$ 
and $J(H)$ are given in Ref.\cite{Zhang2008}, $g({\vec r})$ is the exchange
interaction and $N_2=N/2$ is the particle number in
hyperfine-Zeeman state $\ket{2}$. Again $C_g(a_s)$ is independent of
temperature $T$ and we conclude that the average RF-shift has the same
temperature dependence as the interaction energy at arbitrary
$\xi$. Note that $\overline{\d\o}$ scales with $k_F$ at unitarity.

\subsection{Closed Channel Fraction}
One of the key physical quantities in the BEC-BCS crossover using
Feshbach Resonance is the population in the 
closed channel. This quantity has been experimentally determined using
an optical molecular spectroscopic technique\cite{Partridge2005}. The
experiment uses $^6$Li atoms in their lowest two hyperfine states
$\ket{1}$ and $\ket{2}$, in which they interact primarily through the
electronic
triplet potential. On the other hand, the associated closed channel
molecules induced by the Feshbach resonance interact primarily through
the much deeper electronic singlet potential. In the experiment, a laser
beam induces an electric dipole transition between the closed channel
molecular state($X^1\S_g^+,\nu=38$) to another closed channel
molecular state
with $\nu=68, A^1\S_u^+$. At low temperature, it is inferred from the
loss signal that on the BCS side 
of the resonance, there is a finite fraction of closed channel
molecules which is not supported by the two-body physics. Thus it is
suggested that the many-body quantum state must have non-trivial
two-particle correlations like those in the BCS state to account for the
observed one-body decay in the BCS side\cite{Partridge2005}. To
address the closed channel 
fraction theoretically\cite{Chen2005,Romans2005,Javanainen2005}, let us
first identify the inter-channel coupling 
$W({\vec r})$ from the bare interactions between the two atoms, $U({\vec
r})=f({\vec r})+g({\vec r}_1)\vec{S}_1\cdot \vec{S}_2$.  $f({\vec r})$
and $g({\vec r})$ are the direct and exchange
interaction respectively. We restrict ourselves to the case when there
are only two channels involved, namely, an open channel with atoms in the
lowest two 
hyperfine-Zeeman states $\ket{1}$ and $\ket{2}$ and the corresponding
closed channel with atoms in 
hyperfine-Zeeman states $\ket{1}$ and $\ket{4}$. Notice that one of
the hyperfine-Zeeman states is common to the open and closed
channels. We shall denote the interaction potential in the open and
closed channel by $V_o({\vec r})=\bra{12}U\ket{12}$ and $V_c({\vec
  r})=\bra{14}U\ket{14}$ respectively. Now, the inter-channel coupling
can be written as, 
\ba
W({\vec r}_1-{\vec r}_2)=g({\vec r}_1-{\vec r}_2)\bra{14}S_1\cdot
S_2\ket{12},
\ea
where $\ket{\a\b}$ denotes a spin-singlet state
$\ket{\a\b}=\bk{\ket{\a}_1\ket{\b}_2-\ket{\b}_1\ket{\a}_2}/\sqrt{2}$. The
Hamiltonian is given by,
\begin{widetext}
\ba
\mathcal{H}=\sum_\a\int d{\vec r} \psi_\a^\dag({\vec
  r})\bk{-\frac{\hbar^2}{2m}\nabla^2-\mu_\a+E_\a}\psi_\a({\vec
  r})+\frac{1}{2}\sum_{\a\b\g\d}\int d{\vec r}_1d{\vec
  r}_2\psi_\a^\dag({\vec 
  r}_1)\psi_\b^\dag({\vec r}_2)U_{\a\b\g\d}({\vec r}_1-{\vec
  r}_2)\psi_\g({\vec r}_2)\psi_\d({\vec r}_1) 
\ea
\end{widetext}
where $\mu_\a$ is the chemical potential of the
$\a$-component and $U_{\a\b\g\d}({\vec r})=f({\vec
  r})\d_{\a\d}\d_{\b\g}+g({\vec r})\bra{\a}\vec{S}_1\ket{\d}\cdot
\bra{\b}\vec{S}_2\ket{\g}$. $E_\a$ is the energy of 
hyperfine-Zeeman state $\ket{\a}$. Note that to the extent that the
particle number in any one 
hyperfine-Zeeman level is conserved in the absence of the laser beam,
that is, if we neglect any decay 
of atoms from one hyperfine-Zeeman state to another, we have only two
independent chemical potentials, $\mu_1$ and $\mu_2=\mu_4$,
corresponding to the two separately conserved quantities $N_1=N/2$ and
$N_2+N_4=N/2$.   
To address the population of the closed channel in a many-body system,
we look at the equation of motion for a product of two Fermi
operators, $\psi_\a({\vec r}_1,t)\psi_\b({\vec r}_2,t)$.
\ba\label{twohole}
&&i\hbar\PD{(\psi_\a({\vec r}_1,t)\psi_\b({\vec
    r}_2,t))}{t}\\\nn
&=&\bk{-\frac{\nabla_1^2}{2m}-\mu_\a+E_\a-\frac{\nabla_2^2}
  {2m}-\mu_\b+E_\b}\psi_\a({\vec r}_1)\psi_\b({\vec r}_2)\\\nn
&+&\sum_{\g\d}U_{\a\b \d\g}({\vec r}_1-{\vec r}_2)\psi_\g({\vec
  r}_1)\psi_\d({\vec r}_2)\\\nn
&+&\sum_{\b'\g\d}\int d^3 {\vec r}' U_{\a
  \b'\g\d}({\vec r}_1-{\vec r}')\psi_{\b'}^\dag({\vec r}')\psi_\g({\vec
  r}')\psi_\d({\vec r}_1)\psi_\b({\vec r}_2)\\\nn
&+&\sum_{\b'\g\d}\int d^3{\vec r}' U_{\b
  \b'\g\d}({\vec r}_2-{\vec r}')\psi_{\b'}^\dag({\vec r}')\psi_\g({\vec
  r}')\psi_\a({\vec r}_1)\psi_\d({\vec r}_2).
\ea
Note that since both $f({\vec r})$ and $g({\vec r})$
are short-range functions of order $r_0$, it is clear from
Eqn.(\ref{twohole}) that the conversion from an open channel pair state
to a closed channel molecular state occurs only at short distance, {\it
  i.e.}, $|{\vec r}_1-{\vec r}_2|\sim r_0$. It follows then the last
two terms in 
Eqn.(\ref{twohole}) are of minor importance as compared with the other
terms since they involve another coordinate ${\vec r}'$ which should be
close to ${\vec r}_1$ or ${\vec r}_2$ and thus bring up extra factors of
$k_Fr_0$. They provide either an effective background 
potential or introduce pair states other than the ones under
consideration ($\ket{12}$ and $\ket{14}$) which are relatively
unimportant and thus not of concern here.  In the following, we shall
neglect the last two terms 
in Eqn.(\ref{twohole}). Now, taking Eqn.(\ref{twohole}) to act on the
ground state or thermal ensemble, we find the coupled  equation of
motion of a state with two holes in 
it. Let us denote this state by $\phi_o({\vec r}_1,{\vec r}_2)$ and
$\phi_c({\vec r}_1,{\vec r}_2)$, where subscript o refers to $\a=1,\b=2$
of the open channel and c refers to $\a=1,\b=4$ of the closed
channel. We assume that the rotational degrees of freedom of the
closed channel molecules are not excited at low temperature and remain
in a relative s-wave state. In that case, since $W({\vec r})$ is in fact
isotropic in space, only the s-wave components of
$\phi_o(\vec{r}_1,\vec{r}_2)$ are important in discussing the
population of the closed channel molecules. On the other hand, only s-wave pair
states in the open channel can be converted by a short-range potential
$W({\vec r})$ to closed channel molecules, as is clear from the
structure of Eq.(\ref{2c}). By 
performing a Fourier transform with respect to the center of 
mass coordinate $2{\vec R}={\vec r}_1+{\vec r}_2$ and time, we find the
following coupled equation,
\begin{widetext}
\ba\label{2c}
\bk{\o+\frac{\nabla^2}{m}+\mu_1+\mu_2-E_{\vec K}-V_o({\vec
    r})}\phi_o({\vec r};{\vec K},\o)&=&W({\vec r})\phi_c({\vec r};{\vec
  K},\o)\\\nn 
\bk{\o+\frac{\nabla^2}{m}+\mu_1+\mu_4-E_{\vec
    K}-\tilde{\d}_c-\e_0-V_c({\vec r})}\phi_c({\vec r};{\vec
  K},\o)&=&W({\vec r})\phi_o({\vec r};{\vec K},\o). 
\ea 
\end{widetext}
Here $-\e_0$ is the energy of the molecular state in the closed
channel relative to its 
asymptotic energy $E_2+E_4$ when the two atoms are far away from each
other, $E_{\vec K}=\hbar^2{\vec K}^2/4m$ is the center of mass kinetic
energy of a pair  
of atoms, and $\tilde{\d}_c=E_4-E_2-\e_0$ is the so-called bare
detuning from the 
Feshbach resonance. In case of a broad Feshbach resonance, it is much
larger than the many-body energy scale, in particular, $\tilde{\d}_c\gg
\e_F$. Even though the form of the coupled equation (\ref{2c}) is the
same as that for the two-body case, the many-body physics does play an
important role as it determines the normalizations for the function
$\phi_o(\vec{r}_1,\vec{r}_2)$ and $\phi_c(\vec{r}_1,\vec{r}_2)$. Let
us note one feature of Eq.(\ref{2c}) which is conceptually 
important: Since the inter-channel coupling depends only on the
relative coordinate, so that the center of mass momentum ${\vec K}$
is a good quantum number, we conclude that the pair distributions in
the open and closed channel are connected by Eqn.(\ref{2c}) due to the
superposition nature of the open channel pairs and closed channel
molecules. It is thus in general {\it not} permissible to assign {\it
  independent} momentum distributions to the closed
channel molecules and open 
channel pairs; specifying either one of them suffices to fix the other
through Eqn.(\ref{2c}). Also note that if we
neglect, as we shall do later, the relatively unimportant factor
$E_{\vec{K}}$ as compared 
with $\tilde{\d}_c$, the coupled equation (\ref{2c}) is identical for
different $\vec{K}$-states. That implies that, whatever the center of mass
momentum $\vec{K}$ is for the open channel pair state, the
inter-channel coupling always induces the same amount of closed
channel molecules associated with it. The irrelevance of finite
$\vec{K}$-states in discussing closed channel molecule formation can
again be understood as a result of its high energy character, namely,
the process occurs only at short distance, of order $r_0$ and therefore,
many-body physics is quite incapable of modifying it.


Let us then introduce the Green function for the closed channel
equation.
\ba
\bk{\o+\frac{1}{m}\frac{d^2}{dr^2}-V_c(r)}G(r,r')=-\d(r-r'),
\ea 
where $G(r,r')$ is given by,
\ba\label{gr}
G(r,r')=\sum_n\frac{\chi_n^*(r)\chi_n(r')}{\o-E_n}\approx\frac{\chi_0^*(r)
  \chi_0(r')}{\o+\e_0}.  
\ea
where $\chi_0(r)$ is the normalized eigenfunction in the closed channel
with energy $-\e_0$. Use Eqn.({\ref{gr}}) to integrate the closed
channel equation, we find,
\ba
\phi_c(r;{\vec K},\o)&=&\frac{1}{\o+\mu_1+\mu_4-E_{\vec
    K}-\tilde{\d}_c}\\\nn 
&\times&\int dr'\chi_0^*(r)\chi_0(r')W(r')\phi_o(r';{\vec K},\o). 
\ea
Again, we see that since $W(r')$ is short-ranged, the integration
only picks up the short-range part of the pair wave function. An
important question is the appropriate value for $\o$. It is clear that
the state obtained by removing two particles in states $\ket{1}$ and
$\ket{2}$ does not in general correspond to the eigenstate(or thermal
equilibrium) for the $N-2$ particle system. It is however, clear
that $\o$ will be centered around $-(\mu_1+\mu_2)$, corresponding to
the energy difference between the ground states for the $N-2$- and
$N$-particle 
state. The spread of $\o$ will be in general smaller than the Fermi
energy even at resonance. In the
case of a wide resonance, it is known that $\tilde{\d}_c$ is much larger
than the many-body scale, so if we approximate the denominator in the
above equation by $\tilde{\d}_c$ and make a Fourier transform with
respect to $\vec{K}$, we find 
\ba
\phi_c(r,\vec{R})=-\frac{1}{\tilde{\d}_c}\int dr'
\chi_0^*(r)\chi_0(r')W(r')\phi_o(r',\vec{R}). 
\ea
This implies that the number of molecules in the closed channel $N_C$
is given by, 
\ba\label{nc}
N_C&=&\int dr d\vec{R}\phi_c(r,\vec{R})^*\phi_c(r,\vec{R})\\\nn
&=&\bk{\frac{1}{\tilde{\d}_c}}^2\int
dR dr' dr''K(r',r'')\phi_o(r',\vec{R})\phi^*_o(r'',\vec{R})
\ea
where $K(r',r'')=\chi_0(r')\chi_0^*(r'')W(r')W(r'')$. It is clear that
the kernel $K(r',r'')$ is a short range function in both $r'$ and
$r''$, and that 
$\phi_o(r',\vec{R})\phi^*_o(r'',\vec{R})$ corresponds to the s-wave
part of the 
following density matrix, $\lr{\psi_1^\dag({\vec R}+\frac{{\vec 
    r}'}{2})\psi_2^\dag({\vec R}-\frac{{\vec r}'}{2})\psi_2({\vec
  R}-\frac{{\vec r}''}{2})\psi_1({\vec R}+\frac{{\vec r}''}{2})}$. Using
the same decomposition as before, we find,
\ba
f_c\equiv \frac{N_C}{N}=\bk{\frac{1}{\tilde{\d}_c}}^2 k_F C_K(a_s)
h(\xi,\tau), 
\ea 
where we have used Eq.(\ref{defh}) and defined $C_K(a_s)=\int dr'
dr''\overline{\chi}(r')^*K(r',r'')\overline{\chi}(r'')$. Noting that
the dimension 
of $C_K(a_s)$ is given by $[E]^2[L]$, we can define a length scale
$l_c$ by  
\ba
l_c\equiv\frac{C_K(a_s)}{{\tilde{\d}_c}^2}=\frac{1}{{\tilde{\d}_c}^2}\int dr'
dr''\overline{\chi}(r')^*K(r',r'')\overline{\chi}(r'').
\ea
$l_c$ is entirely determined by the two-body physics. We can rewrite
the molecular fraction in the closed channel as, 
\ba\lb{frac}
f_c=k_Fl_c h(\xi,\tau).
\ea 
At unitarity, $f_c$ scales with $k_F$.

In the following, we shall illustrate the above general considerations
in the `naive' BCS-ansatz, properly generalized to include the closed
channel component. Note that one of the spin state($\ket{1}$) is
common to the open and closed channels, 
\ba
\ket{BCS}=\sum_{{\vec k}}\bk{u_{\vec k}+v_{\vec
    k}a^\dag_{{\vec k}1}a^\dag_{-{\vec k}2}+w_{\vec
    k}a^\dag_{{\vec k}1}a^\dag_{-{\vec k}4}}\ket{vac},
\ea
where $u_{\vec k}$, $v_{\vec k}$ and $w_{\vec k}$ are the usual
variational 
parameters, satisfying $|u_{\vec k}|^2+|v_{\vec k}|^2+|w_{\vec
  k}|^2=1$. $a^\dag_{{\vec k}i}$ is the creation operator for particle 
in hyperfine-Zeeman state $\ket{i}$ with momentum $\vec k$. The
corresponding pair wave functions which are relevant in Eq.({\ref{2c}})
are $F^o_{\vec k}=u_{\vec k}v_{\vec k}$ and $F^c_{\vec k}=u_{\vec k}w_{\vec
  k}$. Let us concentrate only on the $\vec{K}=0$ pair state in the
system since it corresponds to {\it macroscopic} occupation in the
BCS state(cf. also discussion after Eq.(\ref{2c})). Denoting the Fourier
transform of $F^o_{\vec k}$ and $F^c_{\vec k}$ as $F^o(\vec r)\equiv
\sum_{\vec k}u_{\vec k}v_{\vec k}e^{i{\vec k}\cdot{\vec r}}$ and
$F^c(\vec r)\equiv \sum_{\vec k}u_{\vec k}w_{\vec k}e^{i{\vec
    k}\cdot{\vec r}}$ and setting $\o=-(\mu_1+\mu_2)$, we find that
the coupled Eqs.(\ref{2c}) take the form,  
\ba\nn
\bk{\frac{\nabla^2}{m}-V_o({\vec
    r})}F^o({\vec r})&=&W({\vec r})F^c({\vec r})\\\nn 
\bk{\frac{\nabla^2}{m}-\tilde{\d}_c-\e_0-V_c({\vec
    r})}F^c({\vec r})&=&W({\vec r})F^o({\vec r}).  
\ea 
This coupled equation is exactly the same form as that in the two-body
case(see for example, Ref.\cite{Leggett2006}). We can follow the
derivation there or more straightforwardly: we can replace
$\phi_o(r,\vec{R})$ with $\O^{-1/2}F^o(r)$, where $\O^{-1/2}$ accounts
for the center of mass 
motion of the pair and $F^o(r)$ is the radial part of the pair wave
function $F^o(\vec{r})$. The spatial dependence of $F^o(\vec{r})$ is given,
within the crossover model\cite{Zhang2008}, by
\ba
F^o(\vec{r})=\frac{m\D}{4\pi\hbar^2}\frac{1-r/a_s}{r}.
\ea
We find that the density of atoms in closed channel, $n_c$, is given
by, 
\ba
n_c=l_c\bk{\frac{m\D}{\sqrt{4\pi}\hbar^2}}^2. 
\ea
Using $n=k_F^3/3\pi^2$, we obtain the fraction of particles in
the closed channel $f_c$, 
\ba\lb{fbec}
f_c\equiv \frac{n_c}{n}=\frac{3\pi}{16}k_F l_c
\bk{\frac{\D}{\e_F}}^2. 
\ea
If we compare the above equation with Eq.(\ref{frac}), we find
$h(\xi,\tau)=3\pi/16(\D/\e_F)^2$. Thus within the
`naive' ansatz, the fraction of particles in the 
closed channel is proportional to $\D^2$ and moreover, at resonance,
scales with $k_F$. In the extreme BEC limit, we know
$\D=4\e_F/\sqrt{3\pi k_F a_s}$ and thus, 
\ba
f_c=\frac{l_c}{a_s},
\ea
independent of many-body physics, as it should be intuitively. 

Before ending this section, let us make contact with the work in the
literature on the problem of the closed channel fraction. In the work
by Javanainen {\it et.al.}\cite{Javanainen2005}, it is assumed that
the Feshbach induced bosons in the closed channel are condensed in the
$\vec{K}=0$ state. This can be regarded as a limiting case of the
calculation by Chen {\it et.al.}\cite{Chen2005} in which Feshbach
molecules are included in a non-zero temperature generalization of the
conventional `naive' ansatz. The conclusions obtained in
\cite{Chen2005} are in agreement with our general analysis. For
example, it is shown in 
Ref.\cite{Chen2005} that the fraction of condensed bosons scales with
$k_F$ at unitarity and within their approximation, the number of
closed channel molecules(named Feshbach molecules in
Ref.\cite{Chen2005}) is proportional to $\D_{sc}^2$(our $\D$ in
Eq.(\ref{fbec}) above), while the 
number of non-condensed molecules is proportional to
$\D_{pg}^2$(Eq.(9) in \cite{Chen2005}). $\D_{pg}$ describes
non-condensed Fermion pairs, which are of course included in the
general definition of the function $h(\xi,\tau)$. Thus, it is clear
that the general structure of the conclusions is the same in both
approaches. However, as emphasized before, as a result of the coupled
nature of Eq.(\ref{2c}), the momentum distribution of the open channel
pair states dictates the momentum distribution of the closed channel
molecules(Feshbach molecules). While this feature is explicit at the
Hamiltonian level of the two-channel model, it is in general not
enforced in the actual calculations(see for example Eq.(94)in
Ref.\cite{Levin2005}). To illustrate the point, let us look at the
unitarity limit at $T=0$ where we know that a fraction of the Fermi pairs
is not condensed; thus the induced closed channel molecules
associated with them will have nonzero momentum, far from being
condensed in the $\vec{K}=0$ state.

\section{conclusions}
By exploiting the diluteness of the ultra-cold Fermi gas, we have shown
that, in considering various physical quantities of the system, it is
possible to lump all the many-body dependence into a single 
universal function $h(\xi,\tau)$. A particular  physical quantity may be
universal, irrespective of microscopic details({\it e.g.}, the form of the
interaction potential), in which case, one should be able to express
it entirely in terms of $h(\xi,\tau)$, as in the case of the average
single particle energy of the system, Eq.(\ref{spe}). Other physical
quantities are {\it not} universal and there are explicit dependences
on the interaction, other than that incorporated in the function
$h(\xi,\tau)$. However, those dependences can be dealt with using only
the two-body physics. In this case, it is possible to show the
{\it universal} 
temperature dependence of the physical quantities. For convenience,
let us summarize these two-type of behavior in the following,\\
\indent
$\a$), Universal dependence on $\xi$ and $\tau$. It is understood that
universal here means that all the interaction and temperature
dependences are captured in one function $h(\xi,\tau)$. The primary
example is the single particle energy of the system, Eq.(\ref{spe}).
Physical quantities that can be directly derived from energy will be
in this category as well, for example, the speed of sound $c$ and the
chemical potential $\mu$ in Eq.(\ref{mu}). \\
\indent
$\b$), Universal temperature dependence. In this case, the physical
quantities will have identical temperature dependence inherited from
$h(\xi,\tau)$. Those physical quantities include the
interaction energy of the system Eq.(\ref{ie}), the average
radio-frequency spectroscopic shift Eq.(\ref{rf}) and the molecular
fraction in the closed channel Eq.(\ref{frac}).

In actual experiment, there is always an external confining potential
which renders the system inhomogeneous. The question of universality
is then more delicate. However, the argument given in Sec.II is still
valid provided the scale over which the confining potential varies is
much larger than the range of the potential. This is well satisfied in
the experiments. The universal function $h(\xi(\vec r),\tau)$ will
depend on position $\vec{r}$ through local Fermi vector $k_F(\vec
r)$. The temperature dependence of the physical quantities listed in
category $\b)$ above will still have universal temperature dependence
even in a trap.

This work was supported by the National Science Foundation under Grant
No. NSF-DMR-03-50842. 

{\it Note added}. In the process of writing this paper, we become
aware of the recent work of Werner {\it et.al.}\cite{Werner2008},
where an analysis similar in spirit is carried out for the closed
channel molecule fraction. Of particular interest is their definition
of a universal function  which is identical to our $h(\xi,\tau)$, see 
their equation (14). As compared with the paper by Werner {\it
  et.al.}, we have made an effort to connect different physical
quantities together and emphasized the universal temperature
dependences of the physical quantities. In addition, we have discussed
in detail the physical origin of the universal function. See however
in their paper for a discussion of the tail of  the momentum
distribution(their section 3.2).

\begin{appendix}
\section{Expansion of ground state energy around unitarity}
In this appendix, we discuss how to establish the linear dependence of
the ground state energy $\e(\xi)$ on $\xi$ around unitarity. Before we
start with the many-body problem, it is instructive to look at the
two-body problem for guidance. For more details, see
Ref.\cite{Leggett2006}  

Let us consider two atoms of mass $m$, interacting via a central potential
$V_\l(r)$ which can be tuned by a parameter $\l$. The
wave function for the relative motion $\chi_\l(r)$ satisfies the
following time-independent Sch\"{o}dinger equation\cite{Leggett2006},
\ba\lb{2bc}
-\frac{\hbar^2}{2m_r}\frac{d^2}{dr^2}\chi_\l(r)+V_\l(r)\chi_\l(r)=E\chi_\l(r), 
\ea
where $m_r=m/2$ is the reduced mass. Note that the normalization of
$\chi_\l(r)$ is arbitrary at the moment. Let us fix this by requiring
that in the region $r\gg r_0$, 
\ba
\chi_\l(r)=1-\frac{r}{a_s}
\ea
where $r_0$ is the range of the two-body potential $V_\l(r)$. Consider
the critical potential $V_{\l_c}$, for which $a_s=\infty$ and denote
the corresponding radial wave function by $\chi_0$. $\chi_0=1$ for
$r\gg r_0$. For zero-energy
scattering, we have,
\ba\lb{2bcc}
-\frac{\hbar^2}{2m_r}\frac{d^2}{dr^2}\chi_0(r)+V_{\l_c}(r)\chi_0(r)=0
\ea  
Now, multiplying Eq.(\ref{2bc}) by $\chi_0$(setting $E=0$ in the
right-hand side as well for zero-energy scattering) and multiplying 
Eq.(\ref{2bcc}) by minus $\chi_\l$,  we find, using Green's theorem and
integrating up to $r_0$,
\ba\nn
&&\frac{d\chi_0(r)}{dr}\chi_\l(r)-\chi_0(r)\frac{d\chi_\l(r)}{dr}|_{r_0}\\
&=&-\frac{m}{\hbar^2}\int_0^\infty
dr\bk{V_\l(r)-V_{\l_c}(r)}\chi_0(r)\chi_\l(r).  
\ea
Since both $V_\l(r)$ and $V_{\l_c}(r)$ are short-range functions, we
can safely replace $\chi_\l(r)$ with $\chi_0(r)$ around resonance
since they are identical for $r\lesssim r_0$. Using the explicit form
of $\chi_\l(r)$ and $\chi_0(r)$ we find, for infinitesimal change of
$\l$, 
\ba
\d a_s^{-1}=-\frac{m}{\hbar^2}\bk{\int_0^\infty
dr\bk{\PD{V_{\l}(r)}{\l}}_{\l=\l_c}|\chi_0(r)|^2}\d\l 
\ea

Now, let us consider the many-body case. We shall be interested in a
system with $N/2$ spin up atoms with coordinate denoted by
$\vec{x}_i$ and $N/2$ spin down atoms with
coordinate denoted by $\vec{y}_i$, $i=1,2,\cdots,N/2$. The interaction
between spin up atom $i$ and spin down atom $j$ takes the form,
\ba
V_\l(|\vec{x}_i-\vec{y}_j|)
\ea
where $\l$ is a tuning parameter as in the two-body case. It
determines the asymptotic behavior of the many-body wave function in
the range $r_0\ll r \lesssim a_s,k_F$. We denote the corresponding
spatial many-body wave function as $\Psi_\l(\vec{x}_1,\vec{x}_2,\cdots
\vec{y}_1,\vec{y}_2\cdots)$. In general, one is not allowed to write
down a pure spatial wave function with spin part totally decoupled
from it. However, since we are only interested in the
energetics of the system, for which the spin index is only a
spectator, we shall not write the spin component
explicitly. As in the two-body case, the many-body wave function can
be normalized in such a way that for $r_0\ll
|\vec{x}_i-\vec{y}_j|\lesssim a_s,k_F$,
\ba\lb{Aas}
&&\lim_{r_0\ll |\vec{x}_i-\vec{y}_j|\lesssim
  a_s,k_F^{-1}}\Psi(\vec{x}_1,\vec{x}_2,\cdots
\vec{y}_1,\vec{y}_2\cdots)\\\nn
&=&\frac{1}{\sqrt{4\pi}}\frac{1}{|\vec{x}_i-\vec{y}_j|}\bk{1-\frac{|\vec{x}_i
    -\vec{y}_j|}{a_s}}\Psi' 
\ea
where $\Psi'$ is a function of variables other than $\vec{x}_i$ and
$\vec{y}_j$. We {\it say} that the many-body system is on resonance if
$a_s\to \infty$. $a_s$ 
should be regarded as a parameter of the theory which can be tuned in
the experiments by the external magnetic field. Numerically, the value of
$a_s$ as defined in Eq.(\ref{Aas}) must be essentially equal to that
in the two-body case in the same magnetic field, as the discussions in 
Sec.II would imply. 

To put our system in a finite volume such that the density of
particles $n=N/\O$ is kept constant, we shall introduce the
characteristic function of the volume $\O$,
\ba
\chi_\O(\vec{x})=1, \mbox{if $\vec{x}\in \O$, otherwise zero},
\ea
then we can enforce the condition of constant density through an
external one-body potential $U(\vec{x})$,
\ba
U(\vec{x})=U_0(1-\chi_\O(\vec{x}))
\ea
where $U_0$ is a large constant representing the hard wall such the
many-body wave function vanishes outside the region $\O$ and on the
boundary of $\O$, denoted by $\partial\O$. 
\ba
\Psi(\vec{x},\vec{y})|_{\partial\O}=0.
\ea
$\Psi(\vec{x},\vec{y})$ is a short-hand for 
$\Psi(\vec{x}_1,\vec{x}_2,\cdots \vec{y}_1,\vec{y}_2,\cdots)$. The
time-independent Sch\"{o}dinger equation takes the form,
\begin{widetext}
\ba\lb{a1}
\sum_i\bk{-\frac{\hbar^2}{2m}\nabla_{\vec{x}_i}
-\frac{\hbar^2}{2m}\nabla_{\vec{y}_i}+U(\vec{x}_i)+U(\vec{y}_j)}
\Psi(\vec{x},\vec{y})+\frac{1}{2}\sum_{i,j}V_\l(\vec{x}_i-\vec{y}_j)
\Psi(\vec{x},\vec{y})=E_\l\Psi(\vec{x},\vec{y})    
\ea 
\end{widetext}
In writing the above equation, we have neglected the interaction
potential between parallel spins, corresponding to the Fock energy in
a many-body system. This is certainly negligible as compared with the
Hartree term between anti-parallel spins, which has been incorporated
in the above expression.  We now follow the same recipe developed for
the two-body case. We 
write another equation corresponding to $\l=\l_0$, {\it i.e.}
corresponding to 
$a_s\to \infty$ at resonance. Let us also denote the corresponding
energy by $E_0$ and the wave function by $\Psi_0(\vec{x},\vec{y})$. Then by  
multiplying each equation with $\Psi_\l$ or $\Psi_0$ respectively and
subtracting against each other, we find,
\ba
E_\l-E_0=\frac{1}{2}\sum_{i,j}\frac{\int d\vec{x}
  d\vec{y}\d V_\l(\vec{x}_i-\vec{y}_j)
\Psi_\l\Psi_0}{\int d\vec{x}d\vec{y}\Psi_\l \Psi_0} 
\ea 
here $\d V_\l(\vec{x}_i-\vec{y}_j)=V_\l(\vec{x}_i-\vec{y}_j)
-V_{\l_c}(\vec{x}_i-\vec{y}_j)$. We have used Green's theorem and
the fact that the wave function vanishes at the boundary of the volume
$\O$. In differential form, we have,
\ba
\frac{d E_\l}{d\l}=\frac{1}{2}\sum_{i,j}\frac{\int d\vec{x}
  d\vec{y}\PD{V_\l(\vec{x}_i-\vec{y}_j)}{\l}
\Psi_\l\Psi_0}{\int d\vec{x}d\vec{y}\Psi_\l \Psi_0} 
\ea 
Now, $\l$ determines the scattering length $a_s(\l)$ and thus if we
use $\xi=-(k_Fa_s)^{-1}$, we find, by using Eq.(\ref{asl}),
\ba\lb{cop}
\frac{dE(\xi)}{d\xi}&=&-\frac{d\l}{d
  a_s^{-1}}\frac{k_F}{2}\sum_{i,j}\frac{\int d\vec{x} 
  d\vec{y}\PD{V_\l(\vec{x}_i-\vec{y}_j)}{\l}
|\Psi_0|^2}{\int d\vec{x}d\vec{y}\Psi_\l\Psi_0}\\\nn
&=&-\frac{\hbar^2}{2m} k_F \frac{\sum_{i,j}\int d\vec{x}d\vec{y}
  \PD{V_\l}{\l}|\Psi_0|^2}{\int_0^\infty dr
  \PD{V_\l}{\l}|\chi_0(r)|^2}\frac{1}{\int
  d\vec{x}d\vec{y}\Psi_\l\Psi_0} 
\ea
Notice the resemblance of Eq.(\ref{cop}) with Eq.(\ref{fh2}). In fact
we have merely managed to express the function $h(\xi)$ in terms of
the 
many-body wave function. We have replaced $\Psi_\l$ with $\Psi_0$ in
the numerator since $\PD{V_\l}{\l}$ is a short range function. The
integral involving $\PD{V_\l}{\l}$ in the 
numerator only picks up the short-range contribution from the
probability distribution function $|\Psi_0(\vec{x},\vec{y})|^2$ and
thus apart from an normalization constant which is finite, cancels
off the factor $\int_0^\infty dr \PD{V_\l}{\l}|\chi_0(r)|^2$ in the
denominator. Thus, all the sensitive short-range
dependence has disappeared in the above expression and we are left
with quantities that are independent of short-range complications.
Note now that the factor $\int d\vec{x}d\vec{y}\Psi_\l\Psi_0$
approaches a 
constant as $\xi\to 0$ since it is merely the normalization factor for
the wave function $\Psi_0$. This is why, from a many-body wave
function point of view, the definition of the function $h(\xi,\tau)$
is universal to the dilute Fermi gas system, irrespective of its
short-range potential. The complicated expression one the R.H.S. of
Eq.(\ref{cop}) reduces to a simple combination of normalization
constants and thus remains well-defined for $a_s\to \infty$. We have
thus established the linear dependence of the energy on $\xi$ around
resonance. 

Finally, the extension of the above argument to finite temperature is
straightforward. In the case of thermal equilibrium, we instead
consider a distribution of eigenstates $\ket{n}$ with energy $E_n$,
each of them occurring with probability given by the usual Boltzmann
factor $e^{-E_n/k_BT}$. One can repeat, word by word, the above
derivation and thus we can conclude that at finite temperature, the
$\xi$-dependence of energy away from resonance is linear. This of
course, assumes that the temperature is quite low so that its
effects on the short-range wave function is irrelevant. The argument
can be extended also to a non-equilibrium situation where the
probability of state $\ket{n}$ is given by $p_n$,
$\sum p_n=1$. However, it is practically useless since the
characterization of the system as `away from the resonance' is
ambiguous and one is not likely to obtain any useful conclusions
from the argument.

\end{appendix}


\begin{thebibliography}{99}

\bibitem{Bloch2008}
I. Bloch, J. Dalibard and W. Zwerger, Rev.Mod.Phys. {\bf 80},
885(2008). 

\bibitem{Pita2008}
S. Giorgini, L.P. Pitaevskii and S. Stringari, Rev.Mod.Phys. to be
published.

\bibitem{Ketterle2007}
W. Ketterle and M.W. Zwierlein, {\it Ultracold Fermi Gases},
Proceedings of the International School of Physics "Enrico Fermi",
Course CLXIV, Varenna, 20 - 30 June 2006, edited by M. Inguscio,
W. Ketterle, and C. Salomon(IOS Press, Amsterdam) 2008.

\bibitem{Levin2005}
Q. Chen, J. Stajic, S. Tan and K. Levin,  Physics
Reports {\bf 412}, 1-88 (2005). 

\bibitem{Kohler2006}
T. K\"{o}hler, K. G\'{o}ral and Paul S. Julienne,
Rev.Mod.Phys. {\bf 78}, 1311(2006).

\bibitem{Randeria1995}
M. Randeria, in {\it Bose-Einstein Condensation}, edited by
A. Griffin, D.W. Snokes and S.Stringari. Cambridge University Press,
1995. 

\bibitem{note-dc}
The expression for $\d_c$ as given in Ref.\cite{Leggett2006} is in
terms of the parameter $\kappa$ defined by Eq.(4.A.17) in the cited
reference. To obtain the expression in the text, we only need to
notice that when expanding the detuning around resonance, we find,
$\D B= \kappa (\PD{\d}{B})^{-1}=\kappa \D\mu^{-1}$. Then by using
Eq.(4.A.25), we obtain the result in the text. Note that we have
ignored the magnetic field dependence of $\kappa$. See
Ref.\cite{Leggett2006}, pp163 ff.. 





\bibitem{Eagles1969}
D.M. Eagles, Phys. Rev. {\bf 186}, 456(1969).

\bibitem{Leggett1980}
 A.J. Leggett, in Modern Trends in the Theory of Condensed Matter,
 Proceedings of the XVIth Karpacz Winter School of Theoretical
 Physics, Karpacz, Poland, 1980, Springer-Verlag, pp 13-27.

\bibitem{Nozieres1985}
 P. Nozi\`{e}res and S. Schmitt-Rink, J. Low Temp. Phys. {\bf 59}, 195
 (1985). 



\bibitem{Haussmann2007}
R. Haussmann, W. Rantner, S.Cerrito and W. Zwerger, Phys.Rev.A. {\bf
  75}, 023610(2007).

\bibitem{Leggett2006}
A.J. Leggett, {\it Quantum Liquids}, Oxford University Press, 2006.



\bibitem{Carlson2003}
J.Carlson, S.-Y. Chang, V.R. Pandharipande, and K.E. Schmidt,
Phys.Rev.Lett., {\bf 91}, 050401(2003). 

\bibitem{Astrakharchik2004}
G.E.Astrakharchik, J.Boronat, J.Casulleras and S.Giorgini,
Phys.Rev.Lett. {\bf 93}, 200404(2004).

\bibitem{Chang2005}
S.Y. Chang and V.R. Pandharipande, Phys.Rev.Lett., {\bf 95},
080402(2005).

\bibitem{Burovski2006}
Evgeni Burovski, Nikolay Prokof'ev, Boris Svistunov and Matthias
Troyer, Phys.Rev.Lett., {\bf 96}, 160402(2006); see also Erratum,
{\it ibid.}, {\bf 97}, 239902(2006).

\bibitem{Bulgac2007}
Aurel Bulgac, Joaqu\'{i}n E. Drut and Piotr Magierski, Phys.Rev.Lett.,
{\bf 99}, 120401(2007).

\bibitem{Akkineni2007}
Vamsi K. Akkineni, D.M. Ceperley and Nandini Trivedi, Phys.Rev.B.,
{\bf 76}, 165116(2007).

\bibitem{Gezerlis2008} 
Alexandros Gezerlis and J. Carlson, Phys.Rev.C., {\bf 77},
032801(R)(2008). 


\bibitem{Holland2001}
M.Holland, S.J.J.M.F. Kokkelmans, M.L.Chiofalo and R.Walser,
Phys.Rev.Lett., {\bf 87}, 120406(2001).

\bibitem{Timmermans2001}
E.Timmermans, K.Furuya, P.W.Milonni and A.K.Kerman, Phys.Lett.A, {\bf 
  285}, 228(2001).

\bibitem{Gurarie2007}
V. Gurarie and L. Radzihovisky, Annals of Physics, {\bf 322},
2-119(2007).




\bibitem{Ho2004}
T.L. Ho, Phys.Rev.Lett., {\bf 92}, 090402(2004).

\bibitem{Nikolic2007}
Predrag Nikoli\'{c} and Subir Sachdev, Phys.Rev.A, {\bf 75}, 033608
(2007). 



\bibitem{Abrikosov1975}
A.A.Abrikosov, L.P.Gor'kov and I.E.Dzyaloshinski, {\it Methods of
  Quantum Field Theory in Statistical Physics}, Dover Publication,
1975. 

\bibitem{Huang1957}
K.Huang and C.N.Yang, Phys.Rev., {\bf 105}, 767(1957).

\bibitem{Lee1957}
T.D. Lee and C.N.Yang, Phys.Rev., {\bf 105}, 1119(1957).


\bibitem{Nishida2007}
Y.Nishida and D.T.Son, Phys.Rev.A, {\bf 75}, 063617(2007).

\bibitem{Chen2007}
Jiunn-Wei Chen and E.Nakano, Phys.Rev.A, {\bf 75}, 043620(2007).


\bibitem{Chin2004}
C. Chin, M. Bartenstein, A. Altmeyer, S. Riedl, S. Jochim,
J. Denschlag, and R. Grimm, Science {\bf 305}, 1128(2004).

\bibitem{Kinnunen2004}
J. Kinnunen, M. Rodriguez, P. T\"{o}rm\"{a}, Science {\bf 305},
1131(2004). 

\bibitem{Ohashi2005}
Y. Ohashi, A. Griffin, Phys.Rev.A {\bf 72}, 013601(2005).

\bibitem{He2005}
Y. He, Q. Chen and K. Levin, Phys.Rev.A {\bf 72}, 011602(2005).

\bibitem{Yu2006}
Z.Yu, G. Baym, Phys.Rev.A, {\bf 73},063601(2006).

\bibitem{Baym2007}
G. Baym, C.J. Pethick, Z. Yu and M.W. Zwierlein, Phys.Rev.Lett, {\bf
  99}, 190407(2007).

\bibitem{Punk2007}
M. Punk and W. Zwerger, Phys.Rev.Lett, {\bf 99},170404(2007).

\bibitem{Perali2008}
A. Perali, P. Pieri, G.C. Strinati, Phys.Rev.Lett. {\bf 100}, 010402 (2008). 

\bibitem{Zhang2008}
S.Z. Zhang and A.J. Leggett, Phys.Rev.A, {\bf 77}, 033614(2008).


\bibitem{Partridge2005}
G.B.Partridge, K.E.Strecker, R.I.Kamar, M.W.Jack and R.G.Hulet,
Phy.Rev.Lett., {\bf 95}, 020404(2005).

\bibitem{Chen2005}
Qijin Chen and K.Levin, Phys.Rev.Lett., {\bf 95}, 260406(2005).

\bibitem{Romans2005}
M.W.J. Romans and H.T.C. Stoof, Phys.Rev.Lett., {\bf 95}, 260407(2005).

\bibitem{Javanainen2005}
J. Javanainen, M. Ko\v strun, M. Mackie and A. Carmichael,
Phys. Rev. Lett., {\bf 95}, 110408(2005). 


\bibitem{Werner2008}
F.Werner, L.Tarruell and Y.Castin, arXiv.0807.0078v1.


\end{thebibliography}
\end{document}